\documentclass{aa}

\usepackage{graphicx}
\usepackage{subfigure}
\usepackage{balance}
\usepackage{placeins}
\usepackage{float}
\usepackage{txfonts}
\usepackage[colorlinks=true, allcolors=blue]{hyperref}

\begin{document}

   \title{The substellar population in Corona Australis}

   \author{K. Mužić
          \inst{1,2},
          V. Almendros-Abad\inst{3},
          A. Baptista\inst{2},
          A. Scholz\inst{4},
          D. Capela\inst{4},
          S. Pearson\inst{5},
          B. Damian\inst{4},\\
          A. doBrito-doVale\inst{1,2,6},
          T. Rom\inst{7,8} \&
          R. Jayawardhana\inst{9}
          }

   \institute{Instituto de Astrofísica e Ciências do Espaço, Faculdade de Ciências, Universidade de Lisboa, Ed. C8, Campo Grande, 1749-016 Lisbon, Portugal\\
              \email{kmuzic@fc.ul.pt}
         \and
   Departamento de Física, Faculdade de Ciências, Universidade de Lisboa, Edifício C8, Campo Grande, 1749-016 Lisbon, Portugal
\and
    Istituto Nazionale di Astrofisica (INAF) – Osservatorio Astronomico di Palermo, Piazza del Parlamento 1, 90134 Palermo, Italy 
            \and
    SUPA, School of Physics \& Astronomy, University of St Andrews, North Haugh, St Andrews, KY16 9SS, United Kingdom 
            \and
    European Space Research and Technology Centre (ESTEC), European Space Agency, Postbus 299, NL-2200AG Noordwijk, the Netherlands 
            \and
    Laboratoire d’astrophysique de Bordeaux, Univ. Bordeaux, CNRS, B18N, allée Geoffroy Saint-Hilaire, 33615 Pessac, France
            \and 
    University of Split, Faculty of Science, Ruđera Boškovića 33, 21000 Split, Croatia
         \and
   Univ. Grenoble Alpes, CNRS, IPAG, 38000 Grenoble, France 
\and   
Department of Physics and Astronomy, Johns Hopkins University, 3400 N. Charles Street, Baltimore, MD 21218, USA 
             }

   \date{Received ; accepted }
 
  \abstract
   {The substellar initial mass function (IMF) and the formation mechanisms of brown dwarfs (BDs) remain key open questions in star formation theory. Detailed census and characterization of the IMF in a large number of star-forming regions are essential for constraining these processes.
   }
   {We aim to identify and spectroscopically confirm very low-mass members of the Corona Australis (CrA) star-forming region to refine its substellar census, determine its low-mass IMF, and compare it to other clusters.
}
   {Using deep $I-$band photometry from SuprimeCam/Subaru and data from the VISTA Hemisphere Survey (VHS), we identified low-mass BD candidates in CrA. We subsequently conducted near-infrared spectroscopic follow-up of 173 of these candidates with KMOS/VLT, and also obtained optical spectra for eight kinematic candidate members identified via $Gaia$ data using FLOYDS/LCO.
   }
   {The kinematic candidates observed with optical spectroscopy are confirmed as low-mass stellar members with spectral types M1 to M5. In contrast, all 173 BD candidates observed with KMOS are identified as contaminants. Although the follow-up yielded no new substellar members, it places strong constraints on the number of undetected substellar objects in the region. Combined with literature data, this enables us to derive the substellar IMF, which is consistent with a single power-law slope of $\alpha = 0.95 \pm 0.06$ in the range
   $0.01 - 1$\,M\,$_{\odot}$ or $\alpha = 0.33 \pm 0.19$ in the range $0.01 - 0.1$\,M\,$_{\odot}$. The star-to-BD ratio in CrA is $\sim$2. We also provide updated IMFs and star-to-BD ratios for Lupus~3 and Cha~I from the SONYC survey, reflecting revised distances from  $Gaia$. 
   Finally, we estimate surface densities and median FUV fluxes for six star-forming regions and clusters to characterize their environments and compare their substellar populations as a function of environmental properties.
   }
   {The IMF and star-to-BD ratio show no clear dependence on stellar density or ionizing flux from the massive stars. A combined effect - where one factor enhances and the other suppresses BD formation - also appears unlikely.}


  \titlerunning{Substellar population in CrA}
  \authorrunning{Mužić et al.}

   \maketitle
%
\section{Introduction}
\label{sec:intro}
Surveys conducted in star-forming regions (SFRs) and open clusters have shown that isolated brown dwarfs (BDs), objects with masses below 0.075 M$_\odot$, constitute a substantial fraction of the star-like population in our Galaxy \citep[e.g.][]{muzic19, kirkpatrick24}. This population extends into the planetary-mass regime, including objects with masses below the deuterium-burning threshold of approximately 12\,M$_{\rm Jup}$ \citep{pena12,scholz12b,lodieu18,bouy22,langeveld24}. For the more massive BDs, there is broad agreement in the community that their formation mechanisms are similar to those of stars, as indicated by shared characteristics such as circumstellar disks, outflows, binary properties, and the smooth extension of the initial mass function (IMF) into the substellar regime \citep{luhman12}. 
At the lowest masses (below $\sim$12\,M$_{\rm Jup}$), the population may originate from two distinct formation pathways: direct collapse via cloud fragmentation, like stars, or formation in circumstellar disks followed by dynamical ejection \citep{padoan04,boley12,haworth15,daffern22,scholz22}.

Characterizing the IMF in the substellar regime is key to understanding the efficiency and potential universality of star formation across diverse environments. One of the pioneering surveys that aimed to constrain the substellar IMF in the Solar neighborhood was the Substellar Objects in Nearby Young Clusters (SONYC) project (e.g., \citealt{scholz12a, muzic15}). It delivered an unbiased census of substellar populations in four nearby star-forming regions, revealing an IMF that extends smoothly across the hydrogen-burning limit. In the mass range below 1\,M$_\odot$, the IMF can be approximated by a single power-law, dN/dm\,$\propto$\,m$^{-\alpha}$, with slope values typically between 0.6 and 1. However, the spread in reported slopes is likely influenced by systematic uncertainties in IMF derivation, arising from assumptions regarding distance, age, extinction law, and choice of evolutionary models \citep{scholz13}, as well as differences in the mass ranges over which the IMF is fit \citep[][their Fig. 1]{hennebelle24}. 
Other surveys of nearby star-forming regions \citep{bayo11, alvesdeoliveira12, suarez19}, as well as studies of more distant and often more massive young clusters \citep{muzic17, muzic19, almendros23, gupta24}, have generally found IMF slopes in a similar range as mentioned above. In a comparative analysis of eight young clusters, \citet{damian21} found no compelling evidence for significant environmental dependence of the characteristic mass and the standard deviation of the IMF in the log-normal form. 

In this paper, we present a search for the lowest-mass BDs in the Corona Australis (CrA) SFR, using a methodology that closely follows the framework established in the SONYC project.
At a distance of $\sim$150\,pc, CrA is one of the nearest sites of star formation \citep{neuhauser08, dzib18,zucker20, galli20}. 
The most recent census \citep{galli20,esplin22} contains $\sim$350 high-probability member candidates, distributed over two kinematically distinct populations. The younger of the two (as evidenced by twice as large disk fraction) is the ``on-cloud'' population associated with the dense cores and YSO population of the Coronet cluster, which is targeted in this paper. Its age has been estimated to be between 1 and 5 Myr \citep{wilking97,neuhauser08, sicilia11, galli20, esplin22}.

This paper is structured as follows.
In Section~\ref{sec:observations}, we describe the observations and data reduction, and in Section~\ref{sec:data_analysis} the data analysis, including the candidate selection and details of the spectroscopic follow-up. The properties of the low-mass population of CrA are discussed in
Section~\ref{sec:discussion}, and compared to other clusters in \ref{sec:comparison_regions}. We summarize the main findings in Section~\ref{sec:summary}.

\section{Observations and data reduction}
\label{sec:observations}
\subsection{Imaging}

\begin{figure*}[]
   \sidecaption
    \includegraphics[width=0.68\textwidth]{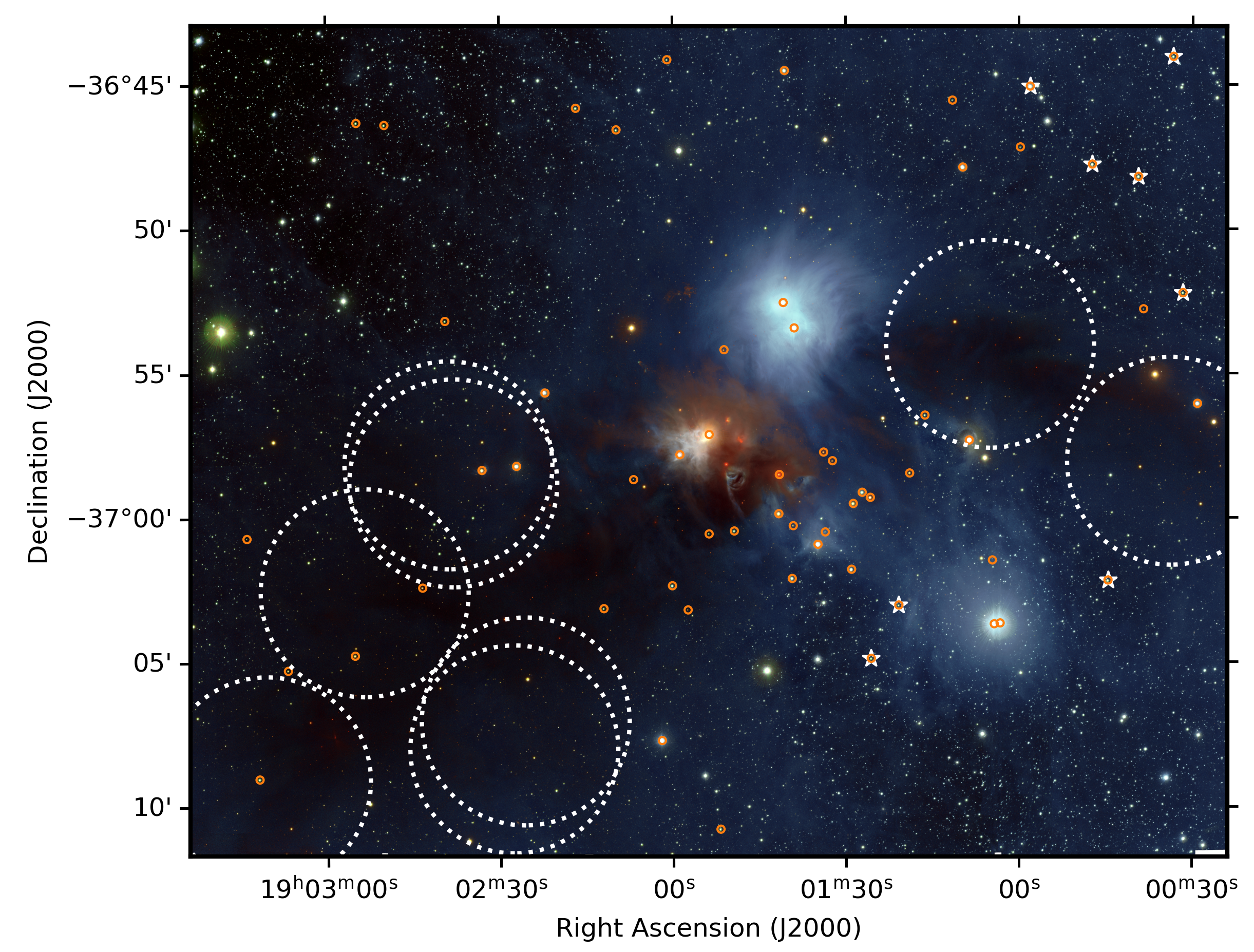}
    \caption{The color composite of the region in Corona Australis studied in this paper. The blue layer is the Suprime-Cam/Subaru $I$-band, while the green and the red layers are $J-$ and $K_S-$band images from the VISIONS survey \citep{visions} downloaded from the ESO science archive. The size of the field is $36.1\times29.1$ arcmin$^2$. The small orange circles indicate members from \citet{esplin22}, while the white stars denote objects observed with FLOYDS/LCO. The dotted white circles mark the fields where targets for follow-up observations with KMOS/VLT were selected.}
    \label{fig:on-sky}
\end{figure*}

\begin{figure}[]
    \centering
    \includegraphics[width=0.45\textwidth]{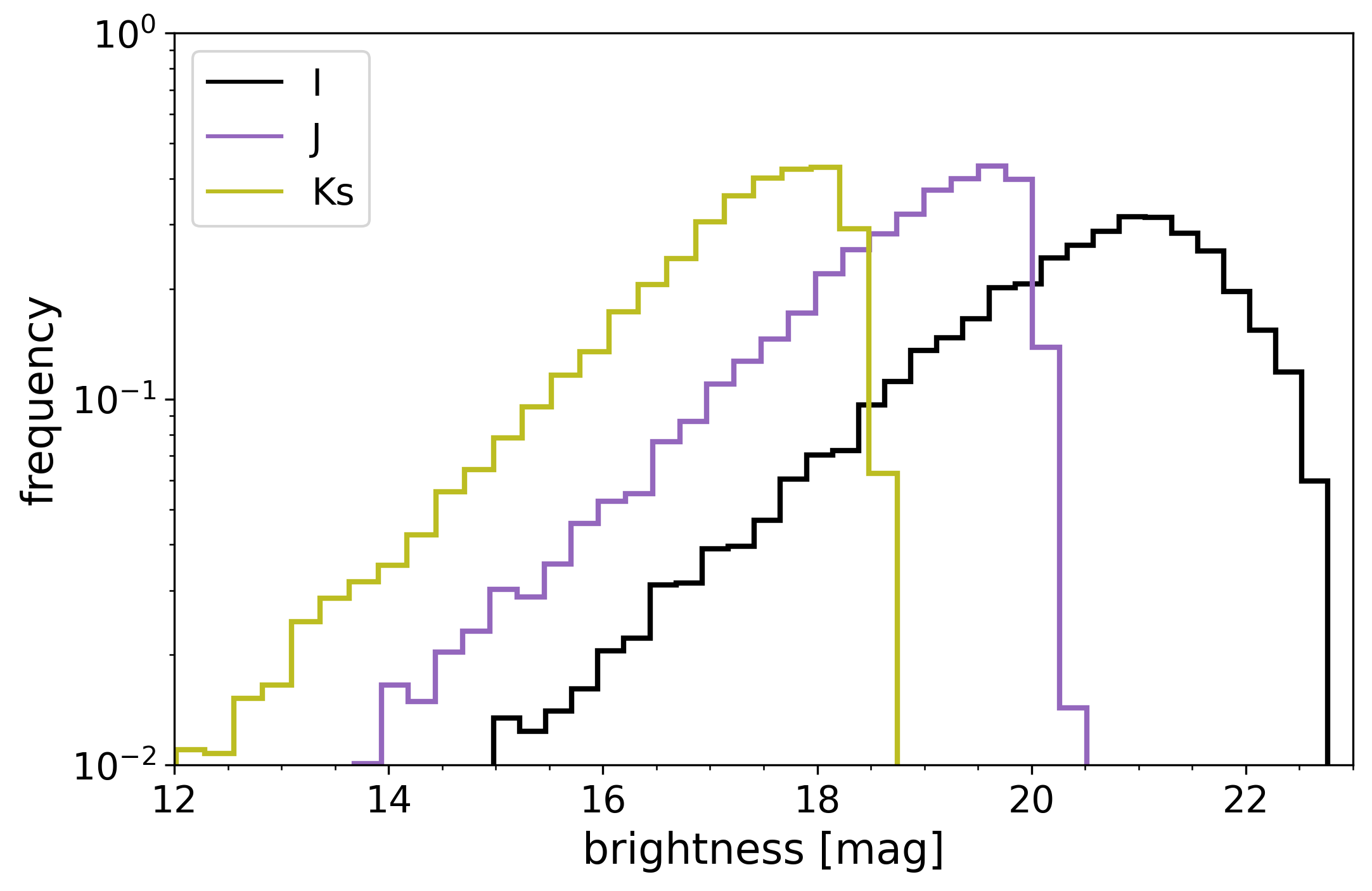}
    \caption{Density of sources in our catalog as a function of magnitude.}
    \label{fig:completeness}
\end{figure}

Images in the $I-$band (filter W-S-I+), taken with Suprime-Cam
at the Subaru Telescope \citep{suprimecam}, were downloaded from the Subaru archive. Suprime-Cam features ten $2048\times4096$ CCDs, covering a total field-of-view of $34' \times 27'$, with a pixel scale of $0.20''$. The total on-source exposure time was 150\,s, split into 5 individual exposures. Data reduction was performed using our own Python routines, including overscan subtraction, flat fielding, bad pixel and cosmic ray correction. The latter was performed using the \textsc{LACosmics} package \citep{lacosmic}. The five images for each detector were aligned using the \textsc{astroalign} package \citep{astroalign} and combined using the median. The combined frames were coordinate-calibrated using \textsc{Astrometry} \citep{astrometry}.
The mean difference between the positions of the sources in our images and their respective values in Gaia DR3 \citep{gaiadr3} is $0.2''$. Finally, a mosaic image was produced using the \textsc{reproject} package \citep{reproject}. The mosaic was used only for visualization purposes, but not for further science analysis which was performed on the individual chip stacks, to avoid differences in background levels due to imperfect scaling between the chips, as well as small-scale distortions observed at the edges of our stacks. 
The region studied in this paper is shown in Fig.~\ref{fig:on-sky}. The blue layer of this color-composite image contains the Suprime-Cam/Subaru $I-$band, while the
green and the red layers represent the $J-$ and $K_S-$band images from the VISIONS survey \citep{visions}.

Source identification and extraction, separately for each CCD, have been performed using the \textsc{Source-Extractor} software package \citep{bertin96}. For object identification, we required at least
5 pixels to be above the 3$\sigma$ threshold detection limit. We further rejected
objects with \textsc{Source-Extractor} keywords FLAGS $\neq 0$, and ELLIPTICITY $\geq$ 0.5.  

We next retrieved $J$ and $K_S$ photometry of the same region from the VISTA Hemisphere Survey (VHS; \citealt{vhs}), and the $I$-band photometry from DENIS survey \citep{denis}. The catalogs were crossmatched with a tolerance of $1''$. The Suprime-Cam photometry was then calibrated using the following expression:
\begin{equation}
    I = I_{instr} + ZP + CT\times(I - J),
\end{equation}
where $I_{instr}$ is the Suprime-Cam instrumental magnitude, ZP is the zero-point, and CT is the color term. Only stars with photometric errors $\leq 0.1\,$mag were taken into account for calibration. The photometric uncertainties were calculated by combining the uncertainties of the zero-point,
color terms, and the measurement uncertainties supplied by \textsc{Source-Extractor}. At the end of the photometry calibration process, all catalogs from the different chips were joined into a single one. For the duplicate sources, we averaged their sky
positions and the photometry. 
To avoid effects of saturation in our catalog, we replaced the $I-$band photometry for those brighter than $I$=16\,mag by DENIS photometry. We also added brighter objects not found in our catalog present in DENIS and VHS.  
Our final catalog contains 19,599 sources. All the sources have $I$ and $J$ photometry, and $72\%$ also contain a $K_S$-band
measurement. 

In Fig.~\ref{fig:completeness} we show the histograms for the three photometric bands and define the completeness limit as the magnitude at which the density of sources reaches a maximum value. We consider our catalog complete down to $I\approx$20.8\,mag, $J\approx$19.5\,mag, and $K_S\approx$18.0\,mag. According to BHAC15 models \citep{baraffe15}, the $J-$ and $K_S$-band limits corresponds to masses <0.01 M$_\odot$ at the distance of 150\,pc, age of 3 Myr, and the extinction of A$_V$=0-5\,mag; for the $I-$band this corresponds to masses from $<$0.01 and up to 0.015 M$_\odot$.

\subsection{Spectroscopic follow-up}

In this section, we describe the observational setup and data reduction for the spectroscopic dataset. The details of target selection for the spectroscopic follow-up will be described in Section~\ref{sec:data_analysis}.

\subsubsection{KMOS/VLT}

The main part of the spectroscopic follow-up has been performed between April and June 2023, using the K-band Multi Object Spectrograph \citep[KMOS,][]{sharpies13} at the Very Large Telescope (VLT) \footnote{ESO program ID 111.24LD (PI: K. Muzic).}. KMOS performs integral field spectroscopy and has 24 arms, each with a square field-of-view (FoV) of $2.8''\times2.8''$. The total FoV of the instrument is $7.2'$ in diameter (white dotted circles in Fig.~\ref{fig:on-sky}). KMOS was operated in service mode using a nod-to-sky ABA configuration, where A is the target and B is the sky. Between each exposure a small dithering was performed. Eight different fields were observed with 10 exposures of 133\,s using the $HK$ filter ($\sim$1.5-2.4 $\mu$m), providing a mean spectral resolving power of $\sim$1800. The following calibrations were also obtained along with the target spectra: darks, flat-field, arc-lamp, and telluric standard stars.

The data reduction was performed using the KMOS pipeline \citep[SPARK,][]{davies13} in the ESO Reflex automated data reduction environment \citep{freudling13}. The pipeline produces calibrated 3D cubes, by performing flat field correction, wavelength calibration, sky subtraction, and telluric correction \citep[using \textit{Molecfit},][]{smette15}. The spectra were extracted by fitting a 2-D Gaussian to the reduced combined cube of each source. In total, 173 source spectra were extracted.

\subsubsection{FLOYDS/LCO}

In June and July of 2021, we obtained optical spectra for 8 objects using the
FLOYDS spectrograph installed at the 2-m Faulkes South telescope, which is part of the Las Cumbres Observatory (LCO) robotic network\footnote{LCO Program ID SUPA2021A-006 (PI: A. Scholz)}. The spectrograph provides a wide wavelength coverage of $3200-10000$\AA~with a spectral resolution R$\sim$400-700, using the $1.6''$ slit. We set an upper wavelength cut-off at 9000\AA~as
a result of the strong telluric absorption present at longer wavelengths. In some cases, we also cut the bluest part of the spectrum if it presents a very low signal-to-noise ratio. Data reduction is identical to that described in \citet{kubiak21}.

\begin{figure*}[htb]
   \center
   \begin{subfigure}{}
        \center
        \includegraphics[width=0.5\linewidth]{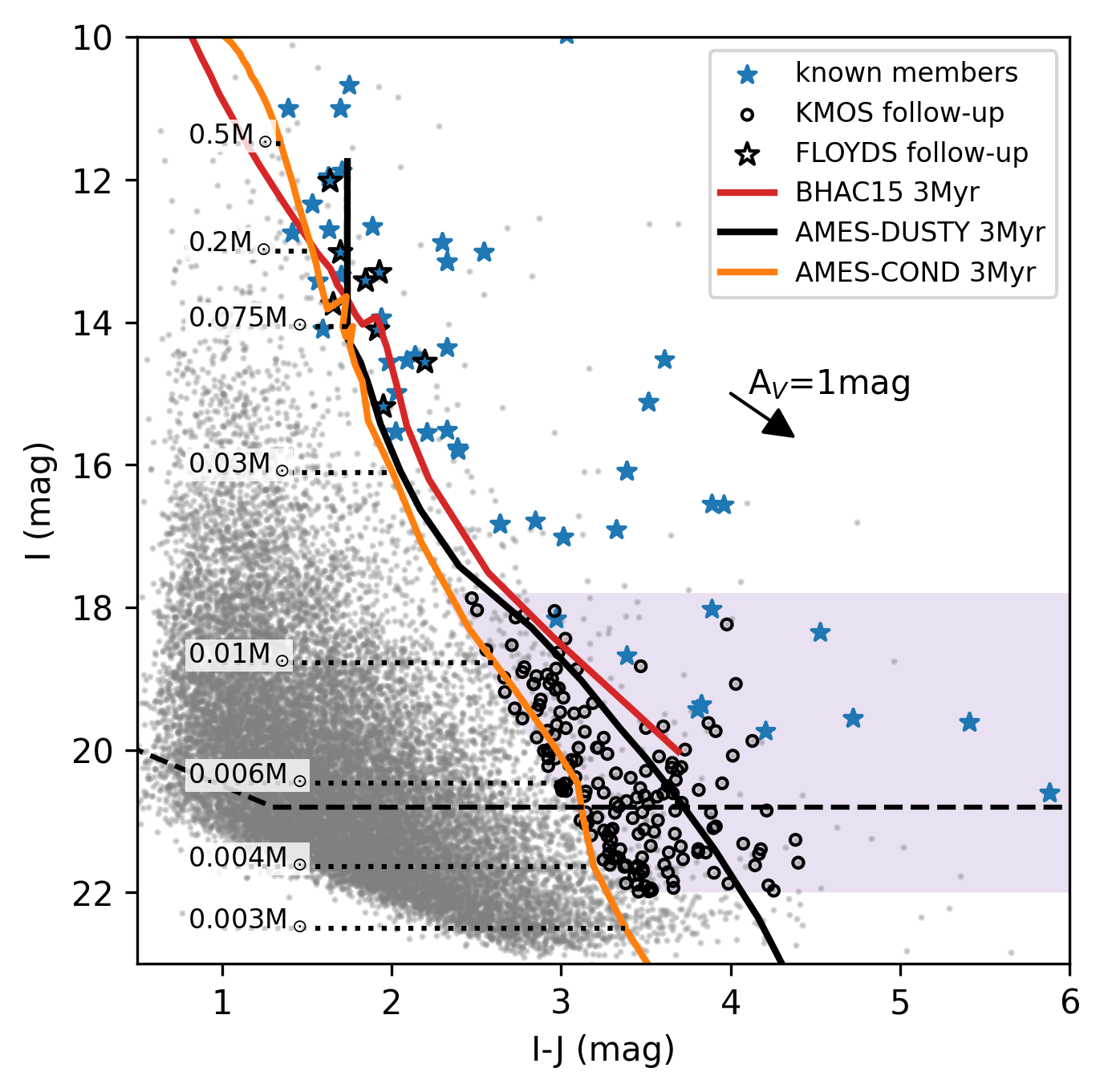}%
        \hfill
        \includegraphics[width=0.5\linewidth]{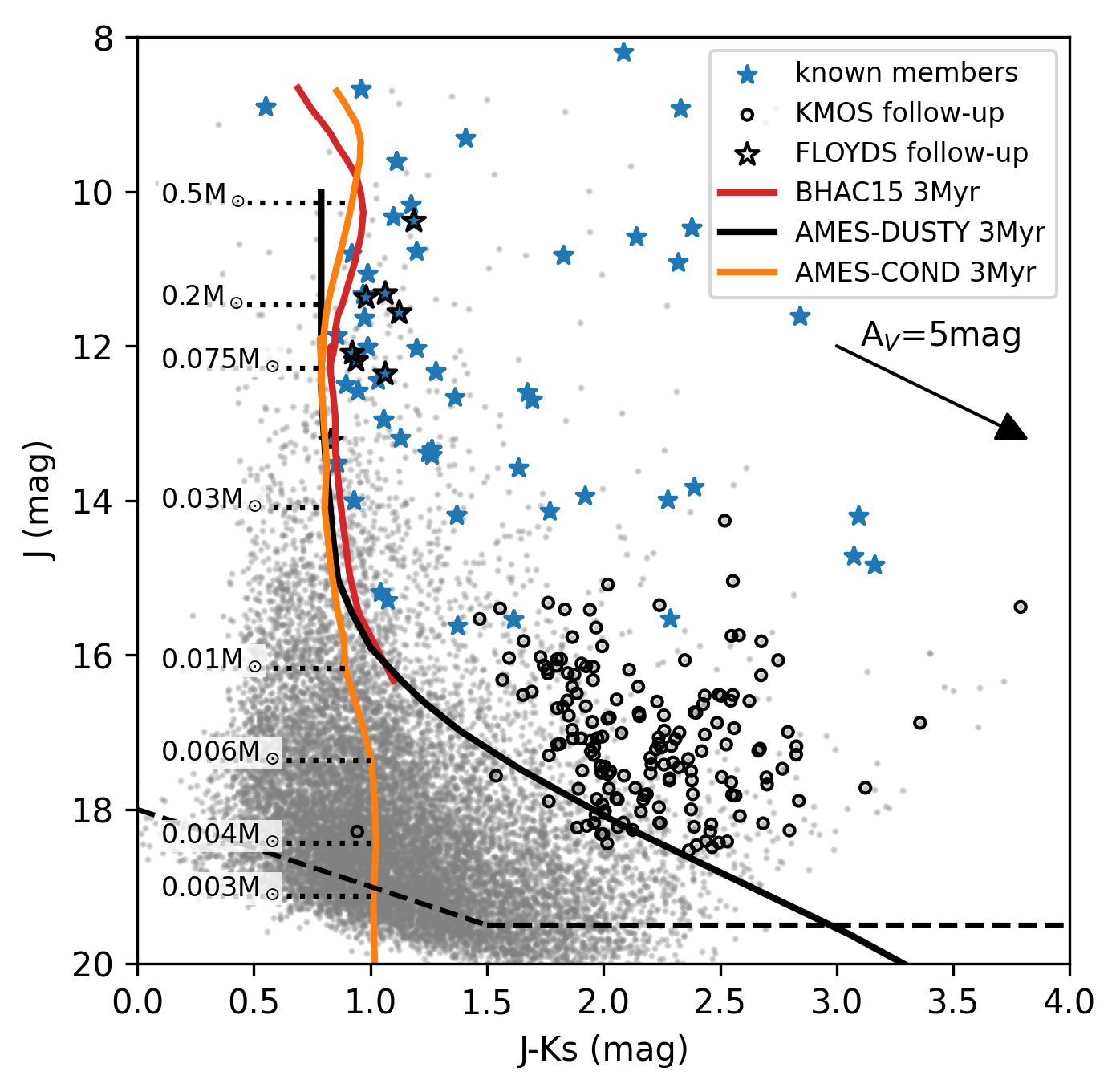}
      \caption{Color–magnitude diagrams showing the sources towards Corona Australis (grey dots), along with the previously confirmed members (blue stars) from astrometry, spectroscopy, MIR-excess or X-ray emission from \citet{peterson11,galli20,esplin22}.
     The sources with spectroscopy obtained in this work are marked with black stars (FLOYDS) and circles (KMOS). The latter were selected from the purple-shaded region. The completeness limit of our catalog is marked with the black dashed line, while the solid lines show the evolutionary models for the age of 3Myr from the AMES-DUSTY (black), AMES-COND (orange), and BHAC15 (red) series \citep{allard01,baraffe15}. For clarity, masses are only marked on the AMES-COND model. The BHAC15 model extends down to 0.01 M$_\odot$.}
         \label{fig:cmd}
    \end{subfigure}
\end{figure*}

\section{Search for low-mass members of CrA}
\label{sec:data_analysis}

\subsection{Candidate selection}

In Fig.~\ref{fig:cmd}, we show the $I$, $I-J$ (left panel) and $J$, $J-K_S$ (right panel) color-magnitude diagrams (CMDs) of the studied region, with grey dots marking all the sources in our catalog, and blue stars showing previously known members selected from Gaia DR2 astrometry \citep{galli20}, mid-infrared excess and X-ray emission \citep{peterson11} and Gaia EDR3 astrometry, along with spectroscopy \citep{esplin22}. The objects included in the spectroscopic follow-up are shown as black open circles (KMOS), and black open stars (FLOYDS). The dashed black lines show the completeness limit, while the solid ones show various 3 Myr isochrones, consistent with the age of CrA. The red is from the BHAC15 model series \citep{baraffe15}, which extends only down to 0.01$\,$M$_{\odot}$. We also show somewhat older AMES-COND and AMES-DUSTY models \citep{allard01}, which were chosen because they extend into the planetary-mass regime. The latter models are available in the DENIS photometric system (important for our $I$-band photometry). In the case of BHAC15 models, we plot their prediction using the Cousins $I$-band, which should be fairly similar to the DENIS $I-$band ($\leq 0.05$\,mag difference; \citealt{delfosse97, costa06,messineo22}). This is further confirmed by cross-matching our catalog with that of \citet{pancino22}, yielding a mean difference of just 0.01\,mag between our $I$-band magnitudes and their $I_C$ values, with a standard deviation of 0.03\,mag.   

The KMOS follow-up was designed to search for the lowest-mass BD and planetary population of the cluster.
The objects for the follow-up were randomly selected from the purple-shaded region shown in Fig.~\ref{fig:cmd}, which was constructed to
extend the sequence of previously known members towards fainter magnitudes, taking the evolutionary models as a guide. Here we decided on the conservative blue-side limit guided by the AMES-COND model. 
The lower edge of the selection box corresponds to the spectroscopic sensitivity of KMOS ($J$=18 mag), while the upper edge was guided by the faint limit of Gaia DR3 ($J\approx15$\,mag; $I\approx17.8$\,mag). While the majority of the candidates fall below Gaia's sensitivity limit, a subset of the sources in the brighter end of the selection box do have Gaia astrometry. An overlap of about 1$\,$mag with Gaia was allowed in order to assure the completeness in the transition region.
The total number of objects in this selection box is 404, of which 173 were observed. None of the observed objects has a spectrum available in the literature. The fields from which the objects were selected are marked by dotted white circles in Fig.~\ref{fig:on-sky}. Their placement was chosen to maximize the number of targets that could be observed simultaneously.

The objects for the follow-up with the FLOYDS spectrograph have been selected from the list of Gaia DR2 kinematic members \citep{galli20}, except the object $\#5$, which was chosen for having a consistent proper motion and parallax from Gaia EDR3. This object, and two more have later been added to the census of CrA by \citet{esplin22} (see Appendix~\ref{sec:galli_rejected}). 

\subsection{Spectral type and extinction}

\subsubsection{KMOS}
\label{sec:kmos}
Based on the candidate selection criteria described in the previous section, if confirmed as members of CrA, majority of KMOS/VLT candidates should have masses between $\sim$4 and 30\,M$_{\rm Jup}$, implying an expected spectral type later than $\sim$M7. Such late-type sources can be easily recognized from their HK band spectra, and their youth can be assessed through the characteristic triangular shape of their $H-$band caused by the effect of strong H$_2$O absorption at both sides \citep{cushing05,allers13,almendros22}.

We start our analysis by visual inspection and find that 
8 candidates show features resembling M-dwarfs. The remaining objects mostly show flat or featureless spectra, and are most probably background stars of spectral type earlier than M, or eventually galaxies. 
For the 8 selected objects, we search for the best-fit template spectrum using a $\chi^2$ minimization:

\begin{equation}
\label{eq:chi2}
    \chi^2=\frac{1}{N-m} \sum_{i=1}^{N} \frac{ (O_i-T_i)^2}{\sigma^2 } , 
\end{equation}

where $O$ is the object spectrum, $T$ the template spectrum, $\sigma$ is the noise of the observed spectrum, $N$ the number of data points, and $m$ the number of fitted parameters ($m$=2). The latter are the SpT, and extinction, which can be fitted simultaneously. 
In the fitting process, we use both young and field spectral templates. The young ones are taken from \citet{luhman17}, and are defined at half-integer SpTs from M0 down to L0, complemented with L2, L4 and L7. The field templates, defined in the range M0–L9 at each 1 subclass interval, are from \citet{kirkpatrick10}. 
Each object is compared with all the spectral templates for a wide range of extinction values: $A_{\mathrm{V}}$=0--20\,mag with a step of 0.5 mag. We use the extinction law from \citet{wang19_extlaw}, which assume $R_{\mathrm{V}}$=3.1. The object spectra were resampled to match the wavelength grid of the templates. 
  The comparison is made for the $H$ and $K$ bands, neglecting the telluric region in between 1.8 and 1.95 $\mu$m. All the spectra were normalized at 1.65 $\mu$m.

The fits confirm that the 8 candidates have the spectral type M, the latest one being M6. In all cases, the field templates provide a better fit.
Additionally, we calculated the $TLI-g$ gravity-sensitive spectral index \citep{almendros22}, and found that 7 are compatible with field dwarfs. The remaining object shows a poor-quality spectrum, and with $I=21.4$\,mag and $J=17.6$\,mag is too faint to be a young mid-M dwarf in the cluster. In conclusion, none of the candidates observed with KMOS are found to be very-low-mass members of the CrA SFR. In Appendix~\ref{sec:kmos_spectra}, we show a selection of the KMOS spectra, and provide coordinates of all the objects included in the KMOS follow-up.

We note that 27 of the objects observed with KMOS have proper motion and parallax data available from Gaia DR3, and none of them are consistent with membership in CrA. 
As mentioned before, some overlap with Gaia was intentional when defining the selection box for the spectroscopic follow-up. However,
 we recognize that with Gaia DR3 available at the time of planning the observations, a more refined selection excluding known kinematic non-members would have improved the efficiency of the KMOS spectroscopic campaign.

\subsubsection{FLOYDS}

\begin{figure*}[]
\sidecaption
    \includegraphics[width=0.7\textwidth]{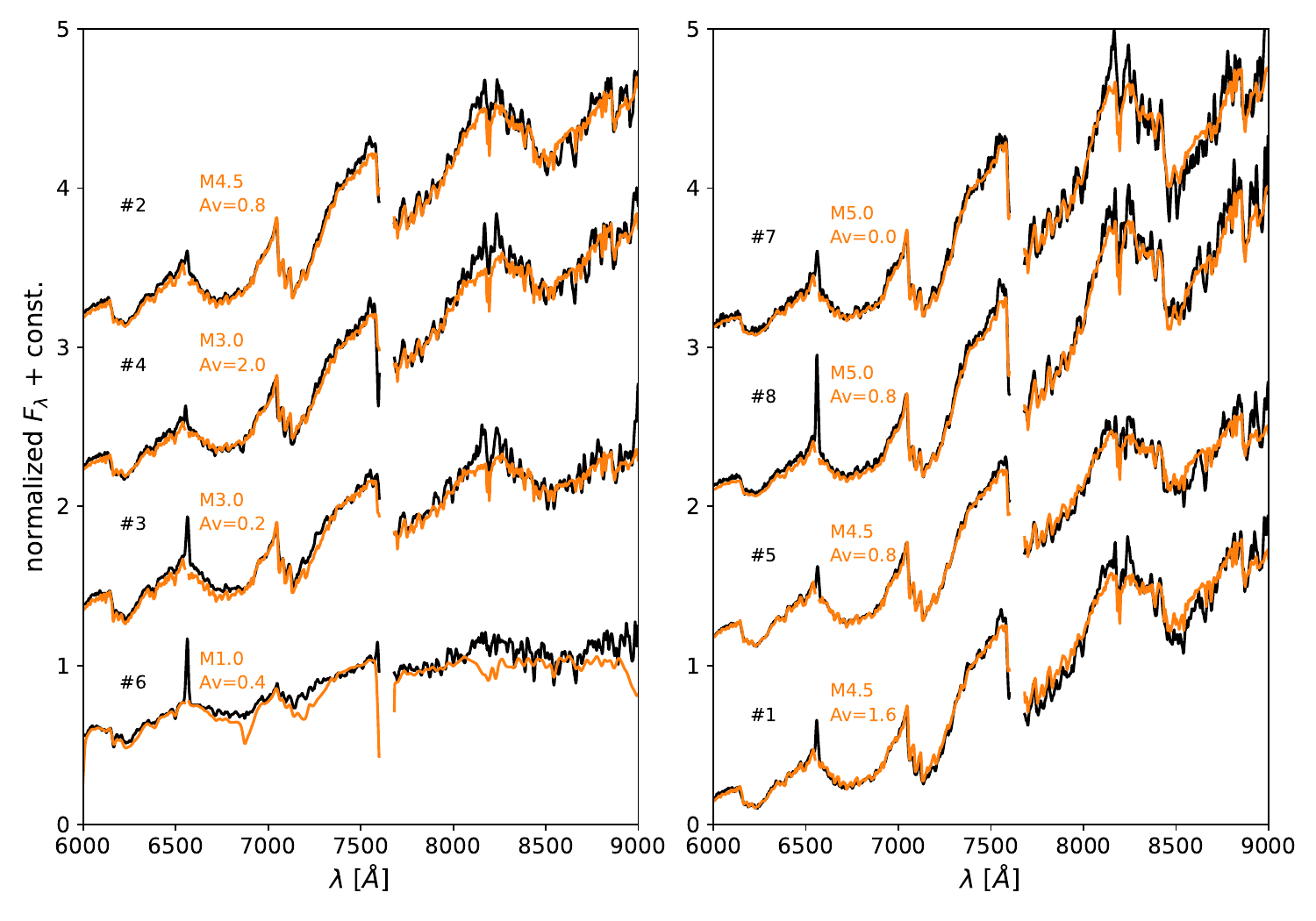}
    \caption{The spectra of the 8 members observed with FLOYDS/LCO (black), along with the best-fit reddened template spectra (orange). All the spectra have been normalized at 7400\AA. The H$\alpha$ emission in the template spectra has been masked for clarity.}
    \label{fig:floyds}
\end{figure*}

\begin{table*}
	\caption{The list of objects followed-up using FLOYDS.}
	\begin{center}
		\begin{tabular}{c c c l c l}
		\hline \hline
		ID & $\alpha$ & $\delta$ & SpT & A$_V$ (mag) & SpT lit. \\
		\hline
		1 & 19:00:31.59	& -36:52:14.0 & M4.5      & 1.6 &  M5$^{a}$ \\
        2 & 19:00:33.27	& -36:44:03.2 & M4.5        & 0.8 &  M4.5$^{a}$   \\
        3 & 19:00:39.30	& -36:48:12.7 & M3     & 0.2 &  M3$^{a}$, M2$^{b}$ \\
        4 & 19:00:44.50	& -37:02:11.2 & M3     & 2.0 &  M3$^{a}$, M2.5$^{c}$\\
        5 & 19:00:47.30	& -36:47:47.1 & M4.5        & 0.8 &  M4.5$^{a}$   \\
        6 & 19:00:58.00	& -36:45:05.3 & M1  & 0.4 &  M0.75$^{d}$, M0.5$^{a}$ \\
        7 & 19:01:20.85	& -37:03:03.2 & M5     & 0.0   &  M5.5$^{a}$, M4.5$^{e}$  \\
        8 & 19:01:25.64	& -37:04:54.0 & M5     & 0.8 &  M5.5$^{a}$, M5.0$^{e}$  \\
			\hline
		\end{tabular}
  		\tablefoot{References: (a) \citet{esplin22}, (b) \citet{walter97}, (c) \citet{sicilia11}, (d) \citet{romero12}, (e) \citet{sicilia08}
	}
	\end{center}
	\label{tab:floyds}
\end{table*}

The follow-up with FLOYDS was executed before the publication of the study by \citet{esplin22}, in which all of the eight objects appear as kinematic members, and were additionally confirmed by spectroscopy. Several objects also have spectra available in earlier studies \citep{walter97,romero12,sicilia08,sicilia11}. 
We derive spectral types and extinction using the $\chi^2$ minimization (eq.~\ref{eq:chi2}), using as templates spectra of young objects (1–10 Myr) from \citet{luhman03a,luhman04a,luhman04b,luhman04c,bayo11,venuti19} that have spectral types M0.5–M9 with a step of 0.25–0.5 spectral subtypes.
The best-fit results are given in Table~\ref{tab:floyds} and shown in Fig.~\ref{fig:floyds}. The derived spectral types are in excellent agreement with those found in the literature. 

\section{Substellar census and the IMF in CrA}
\label{sec:discussion}

\subsection{Estimate of the number of missing CrA members}
\label{sec:missing}

Using the statistics gained from the spectroscopic survey, we can estimate the number of members missing in the current census of CrA. Similar to the works from the SONYC
series, we use the confirmation rates of our spectroscopic follow-up to
estimate the expected number of members among the
remaining photometric candidates. Here, the confirmation rate is defined as the number of confirmed members divided by the
number of candidates with spectroscopy. The uncertainties are represented as 95\% confidence intervals (CI), calculated using the Clopper–Pearson method (suitable for small-number
events; \citealt{gehrels86,brown01}). 

We assume that the census of CrA members is largely complete for sources above our selection area in Fig.\ref{fig:cmd} ($I<17.8$\,mag). The vast majority of objects at these magnitudes have astrometric data from $Gaia$, which has been effectively used to identify candidate members \citep{galli20, esplin22}. The extent of $Gaia$’s coverage in this region of the CMD is also evident in Fig.~\ref{fig:cmd_zoom}.

To estimate the number of potential members still missing in our census, we focus on objects that fall within the selection area (purple-shaded region) and are brighter than the completeness limit (dashed line in Fig.~\ref{fig:cmd}). 
The known members (blue stars in the same figure) are excluded from the analysis. The total number of objects is then 153, of which we took 77 spectra. The confirmation rate is therefore 0/77 = 0$^{+0.047}_{-0.000}$, which means that among the 78 objects without spectra, we may expect 0 to 4 yet undetected members. If we repeat the same analysis, this time removing the objects rejected by Gaia data \citep{galli20} from consideration, we have 78 objects in the selection box, of which 57 were observed with KMOS. This yields a confirmation rate of 0$^{+0.063}_{-0.000}$, and 0 to 1 missing objects.
Furthermore, we may want to consider only the objects to the right of the AMES-DUSTY isochrone. Originally, our selection has been guided by the AMES-COND model, which may be too blue at substellar masses (judging from the sequence of known members). We thus repeat the same analysis as before, keeping only the sources to the right of the AMES-DUSTY model. This results in 0 to 5 missing objects when ignoring the Gaia rejections, and 0 to 3 missing objects taking it into account. 

In summary, our analysis shows that the current census in CrA may be missing at most 5 objects, with masses between 0.006 and 0.07\,M$_\odot$ (estimated in Section~\ref{sec:masses}).

\subsection{Derivation of mass and extinction}
\label{sec:masses}

The masses have been derived for the known members contained in the latest census by \citet{esplin22}, as well as for the objects in our selection box that were not followed-up spectroscopically. 
For the latter, these mass estimates are conditional on the assumption that the object is a true member of Corona Australis, which we know may hold only for a small subset of these objects (Section~\ref{sec:missing}). These mass estimates will be used to correct the IMF for the objects potentially missing in the census.

To estimate the mass and the extinction of an object, we deredden its photometry to the 3-Myr isochrone in the $I$, $I-J$ CMD, which has been shifted to the distance of the cluster given in \citet{galli20}. The uncertainty on the distance is small ($<1$\,pc), and does not have an effect on the mass estimate. We use the extinction law from \citet{wang19_extlaw}. The procedure described here has been repeated for each of the three isochrones shown in Fig~\ref{fig:cmd}. 
The choice of the age is driven by the range of age estimates of CrA in the literature that varies between 1 and 5 Myr \citep{neuhauser08,galli20,esplin22}. At the same time, the disk fraction has been estimated to be around $50\%$ \citep{lopez-mart10,peterson11, esplin22}, which is consistent with an age of 2-3\,Myr \citep{stolte15,richert18,michel21,winter22}. 

To assess how photometric uncertainties impact the determination of extinction and mass, we employ a Monte Carlo approach. For each source, we generate 1,000 synthetic magnitudes per band, drawing from a normal distribution centered on the observed value, with a standard deviation equal to the corresponding photometric uncertainty. For each of these 1,000 realizations, we then derive the mass, and extinction by dereddening the photometry to match the isochrone.
The final A$_V$ value for each source is taken as the average across all realizations, with its uncertainty given by the standard deviation. However, the resulting mass distributions are typically not normally distributed, due to the non-linear relationship between magnitude and mass. To account for this, we retain the full distributions of mass for each source. These are then used to draw individual mass values during the Monte Carlo simulation used to derive the IMF and star-to-BD number ratio.

\subsection{Initial mass function in CrA}
\label{sec:IMF}

\begin{figure*}[]
    \centering
    \includegraphics[width=1\textwidth]{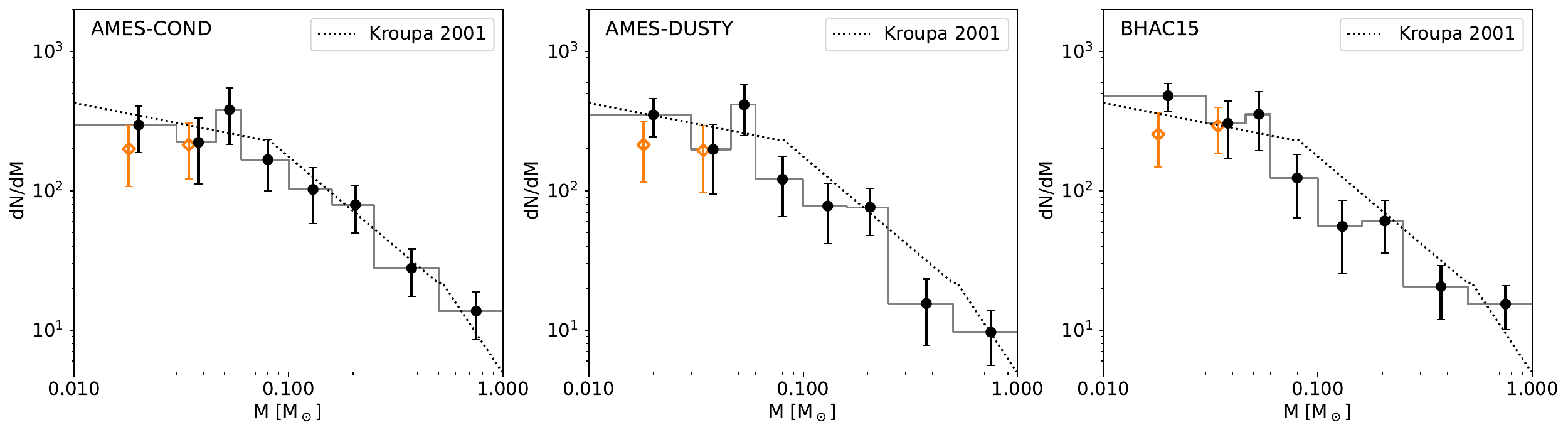}
    \includegraphics[width=1\textwidth]{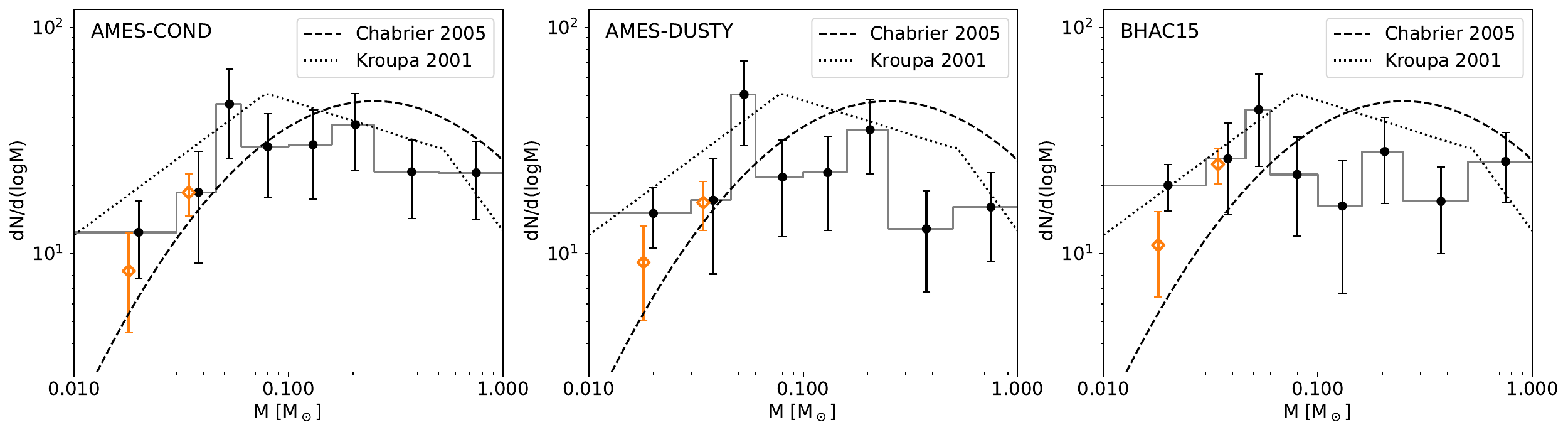}
    \caption{Initial mass function in Corona Australis, in the power-law (top panels) and the log-normal form (bottom panels). 
    The masses were derived
    using three different 3-Myr evolutionary models, indicated in the upper left corner of each panel. The black symbols and the gray histogram represent the IMF after correcting for the missing objects, and the orange points indicate the two affected bins before this correction. The orange points were slightly shifted to the left of the black points, for clarity; the bin centers and sizes were maintained. 
    Two different IMF representations are shown: dN/dm (top panels) and dN/dlog(m) (bottom
panels). The dotted line is the Kroupa segmented power-law mass function \citep{kroupa01}, and the dashed
line shows the Chabrier mass function \citep{chabrier05}, both normalized to match the total number of objects in the cluster. }
    \label{fig:imf}
\end{figure*}

To derive the IMF, we take all the previously confirmed members of CrA located within our survey region from the latest census of \citet{esplin22}. We complement this with the results of our spectroscopic survey, which give a stringent upper limit on the number of members potentially missing within our selection box. 
The stellar masses are sampled from the mass distributions derived in Section~\ref{sec:masses}. This sampling is repeated 100 times for each source, resulting in 100 cluster mass distributions. For 
each of these 100 realizations, we perform 100 bootstrap resamplings — i.e., random selections with replacement. Each bootstrap involves drawing a new sample of N stars (where N is the size of the realization), with replacement, allowing individual stars to be selected multiple times or not at all.
This procedure yields a total of 10,000 mass distributions, which are then used to compute the mass function along with the associated uncertainties.
The IMFs are plotted using a variable size, aiming to maintain the number of objects in each bin roughly constant.

To account for objects potentially missing in the census, we repeated the IMF derivation, randomly selecting 5 objects without spectra from our selection box, and repeating the process 100 times to ensure the statistical robustness of the result. As estimated in Section~\ref{sec:missing}, this is an upper limit on the number of missing objects. The resulting IMFs, in power-law and log-normal forms, are shown in Fig.~\ref{fig:imf}, for three different evolutionary models. The black symbols and the gray histogram represent the IMF after correcting for the missing objects, which affects only the two lowest-mass bins. The orange points indicate the IMF before this correction. For reference, the dotted and dashed lines show the IMFs from \citet{kroupa01} and \citet{chabrier05}, respectively. 
While our overall results are consistent with the former, we note discrepancies with the latter. Specifically, our IMF does not exhibit the sharp decline in the number of substellar objects predicted by the Chabrier functional form. A similar behavior has also been observed in earlier works, e.g. \citet{pena12,muzic19, miret22}.

To enable comparison with previous works, we fit a power-law function to the IMFs shown in the top panels, for three different mass ranges. First, we fit a single slope over the entire available mass range (0.01 - 1\,M$_\odot$). We also provide slopes over the mass range 0.01 - 0.2\,M$_\odot$, and finally in the substellar regime (0.01 - 0.1\,M$_\odot$). By comparing the slope across these different mass ranges, we can observe a gradual flattening of the IMF. The flattening is less pronounced in the case of the BHAC15 model, which may be partially attributed to the fact that this model does not extend below 0.01\,M$_\odot$. As a result, the mass distributions for objects near this limit are truncated at the low-mass end. The fitting was carried out for both the original IMF and the version that includes additional (missing) objects. As the true distribution likely lies between these two cases, we report in Table~\ref{tab:imf} the average value of the slope $\alpha$. For the full mass range (0.01 - 1\,M$_\odot$), we obtain $\alpha$ = $0.95\pm0.06$. In the range 0.01 - 0.2\,M$_\odot$, the slope is $\alpha$ = $0.63\pm0.14$, and in the substellar range (0.01 - 0.1\,M$_\odot$), $\alpha$ = $0.33\pm0.19$.

\begin{table}
\small
	\caption{The slope $\alpha$ of the IMF in Corona Australis in the power-law form and star-to-BD number ratio for three different sets of isochrones.}
	\begin{center}
    \renewcommand{\arraystretch}{1.5}
		\begin{tabular}{l c c c}
		\hline \hline
            \multicolumn{4}{c}{Slope of the IMF ($\alpha$)}\\
            \hline
            &  0.01 - 1 M$_\odot$ & 0.01 - 0.2 M$_\odot$ &  0.01 - 0.1 M$_\odot$\\
             \hline
        AMES-COND & $0.90\pm0.10$ & $0.44\pm0.21$ & $0.15\pm0.28$ \\
        AMES-DUSTY& $1.00\pm0.12$ &  $0.63\pm0.27$ & $0.30\pm0.40$ \\
        BHAC15    & $0.96\pm0.07$ &  $0.83\pm0.23$ & $0.54\pm0.27$\\ 
        mean & $0.95\pm0.06$ & $0.63\pm0.14$ & $0.33\pm0.19$\\
			\hline
            \hline
             \multicolumn{4}{c}{Star-to-BD ratio}\\
            \hline
            Stars (M$_\odot$) & 0.075 - 1 & 0.075-1\\
            BDs (M$_\odot$) & 0.03 - 0.075 & 0.02 - 0.075 \\
               \hline
    AMES-COND & $2.5^{+3.0}_{-1.2}$ &  1.9$^{+1.9}_{-0.8}$\\
    AMES-DUSTY& 1.8$^{+2.5}_{-0.9}$ & 1.3$^{+1.4}_{-0.7}$\\
       BHAC15 & 2.0$^{+2.7}_{-1.0}$ & 1.4$^{+1.4}_{-0.7}$\\  
       \hline
		\end{tabular}
  		\end{center}
	\label{tab:imf}
\end{table}

\subsection{Star-to-brown dwarf ratio}
\label{sec:sbdratio}

\begin{figure*}[]
    \centering
    \includegraphics[width=0.45\textwidth]{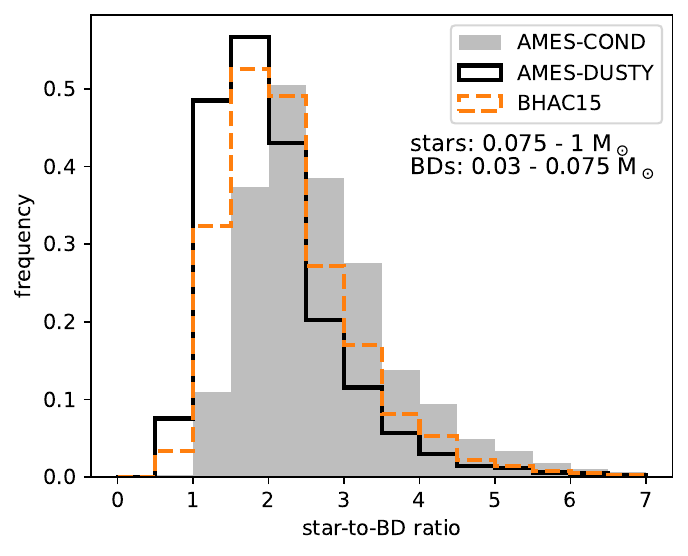}
    \includegraphics[width=0.45\textwidth]{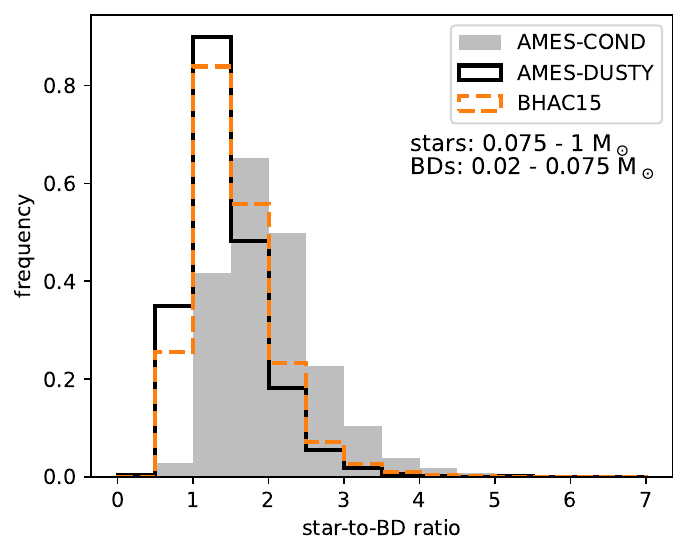}
    \caption{Distribution of star-to-brown dwarf ratio obtained from our data, for three different isochrones. The mass ranges used for the ratio calculation are indicated in each panel. The correction for potentially missing objects in the CrA census was not applied, as their inclusion has little effect on the determination of the star-to-BD ratio.}
    \label{fig:startobdratio}
\end{figure*}

To estimate the ratio between the number of stars and that of BDs, we use the mass estimates derived in Section~\ref{sec:masses} and apply the same Monte Carlo method used in deriving the IMF, generating 10,000 mass distributions. The boundary between stellar and substellar objects is defined at 0.075\,M$_\odot$, corresponding to the hydrogen-burning limit at solar metallicity.
We compute the ratio using two lower mass limits on the BD side: 0.02\,M$_\odot$ and 0.03\,M$_\odot$. For the stellar population, we adopt an upper mass limit of 1\,M$_\odot$, as is commonly done in the literature. 
The resulting distributions of the star-to-BD ratio are shown in Fig.~\ref{fig:startobdratio}, and are also summarized in Table~\ref{tab:imf} for each of the three isochrones considered in this study. The quoted values represent the median of each distribution, and the uncertainties refer to the 95\% confidence limits. We note that the star-to-BD ratio is only minimally impacted by the missing sources; therefore, we report results for the uncorrected case only. The star-to-BD ratio in CrA is $\sim$2, but with a relatively large plausible range, which is mainly the consequence of the small number of members combined with uncertainties in mass determination.

\section{Comparison to other regions}
\label{sec:comparison_regions}
In this section, we compare the IMF and star-to-BD ratio derived for CrA with those from our previous works, which includes three nearby SFRs studies in the SONYC series\footnote{Two of the SONYC IMFs have been recalculated to reflect a significant change in distance since the original publication; the details are given in Appendix~\ref{sec:IMF_updated}.} (Cha I, Lupus 3, and NGC 1333; \citealt{scholz12a,muzic15}), as well as two massive clusters RCW\,38 and NGC\,2244 taken from \citet{muzic17,muzic19}\footnote{In the case of NGC\,2244, we use the IMF for the isochrone $\#1$ and the distance of 1400\,pc, which is close to the latest distance to the cluster determined from $Gaia~DR3$ \citep{muzic22}. A spectroscopic IMF for NGC 2244 was presented in \citet{almendros23}, but it is not included here as it only extends down to 0.045\,M$_\odot$. Nonetheless, it is consistent with the IMF from \citet{muzic19}.}. 

\subsection{IMF comparison}

\begin{figure*}[]
    \centering
    \includegraphics[width=0.8\textwidth]{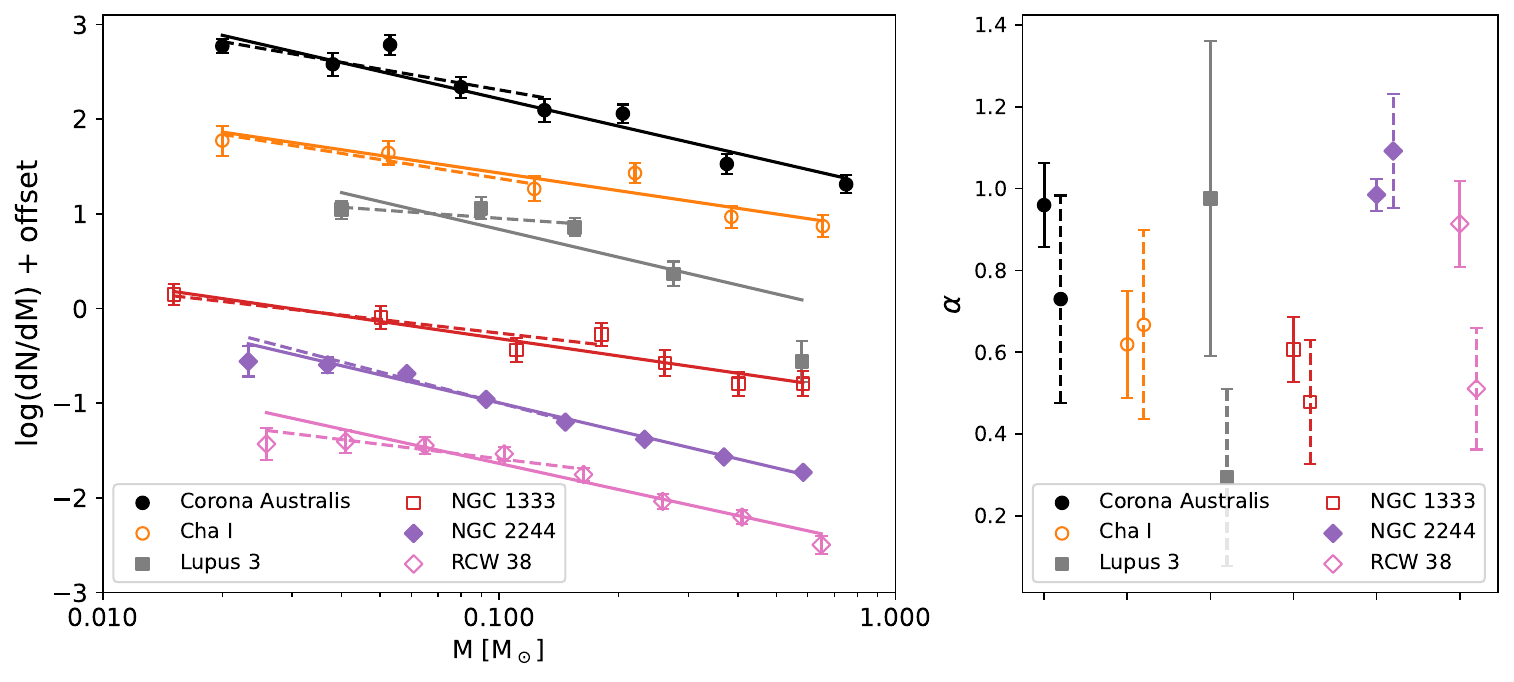}
    \caption{Left: A comparison of the initial mass functions below 1\,M$_\odot$ across six different star-forming regions. The IMF for Corona Australis is the mean of the IMFs presented in Section~\ref{sec:IMF}.  Solid lines represent fits to the full set of data points for each cluster, whereas dashed lines indicate fits restricted to the substellar regime below 0.2\,M$_\odot$.
    Right: The IMF slope $\alpha$ corresponding to the fits displayed in the left panel, indicated using matching linestyles. In Table~\ref{tab:imf_regions}, we list the values shown in this panel, along with the references. }
    \label{fig:imf_comp}
\end{figure*}

\begin{table*}
\small
	\caption{The slope $\alpha$ of the power-laws fitted to the IMF (Fig.~\ref{fig:imf_comp}) and star-to-BD number ratio for different regions.}
	\begin{center}
    \renewcommand{\arraystretch}{1.5}
		\begin{tabular}{l c c c  c c c c c  }
            \hline \hline
                   &     \multicolumn{2}{c}{Slope of the IMF ($\alpha$)} & \multicolumn{2}{c}{Star-to-BD ratio } & reference & surface density$^{d}$ & F$_\mathrm{FUV,median}$\\
            \hline
            &  $<$1M$_\odot$ & $<0.2$M$_\odot$ & 0.03 - 1M$_\odot$  &0.02 - 1M$_\odot$ & &  (pc$^{-2}$) & (G$_0$)\\
            
             \hline
    CrA & $0.95\pm0.06$&  $0.63\pm0.14$ & $2.0^{+2.7}_{-1.0}$ & $1.4^{+1.4}_{-0.7}$  & this work$^{a}$ & 120 & 13.0 \\
    Cha I&  $0.62\pm0.13$&  $0.67\pm0.23$ & $3.2-4.8$& $2.6-3.7$ & \citet{muzic15}$^{b}$ & 29 & 1.2 \\
    Lupus 3    &  $0.98\pm0.38$ & $0.29\pm0.22$ & $2.1-4.5$& $2.0-4.0$ & \citet{muzic15}$^{b}$ & 24 & 0.6 \\ 
    NGC 1333 & $0.61\pm0.08$&  $0.48\pm0.15$ & $1.8-2.8$ & $1.8-2.7$& \citet{scholz12a} & 200 & 27.1 \\
    NGC 2244 &  $0.98\pm0.04$&  $1.09\pm0.14$ & $2.34\pm0.25$& $1.74\pm0.28$ & \citet{muzic19}$^{c}$ & 33 & 11569.8 \\
    RCW 38    &  $0.91\pm0.11$& $0.51\pm0.15$ & $2.0\pm0.6$ & & \citet{muzic17} & 3620 & 266124.6\\ 
       \hline

               \hline
		\end{tabular}
          		\tablefoot{(a) the mean of the results from the three isochrones; (b) updated as described in Appendix~\ref{sec:IMF_updated}; (c) values for the isochrone $\#1$ and distance of 1400\,pc; (d) values from \citet{muzic19} and Appendix~\ref{sec:density}.
	}
  		\end{center}
	\label{tab:imf_regions}
\end{table*}

In Fig.~\ref{fig:imf_comp}, we present the IMFs of CrA and the other five regions listed above.
When considering the full mass range below 1\,M$_\odot$, the values of $\alpha$ range between 0.6 and 1 (see Table~\ref{tab:imf_regions}). The IMF in Corona Australis is on the steeper side of this range, similar to NGC\,2244 and Lupus 3 ($\alpha \sim 1$). A similarly steep slope ($\alpha=1\pm0.2$) has also been derived for Berkeley~59 over the mass range 0.04 - 0.4\,M$_\odot$ \citep{panwar24}. Several IMFs shown in Fig~\ref{fig:imf_comp} show some degree of flattening in the substellar regime. To illustrate this, we also show the fit to the IMF below 0.2\,M$_\odot$ (dashed lines). The flattening is most significant in RCW\,38 (see right panel of Fig.~\ref{fig:imf_comp}) and Lupus~3, however, due to the small number of points in the IMF, the errorbar on the latter result is also quite large. 
For CrA, the flattening becomes more clearly pronounced at masses below 0.1\,M$_\odot$, as discussed in Section~\ref{sec:IMF}.

Flat slopes in the substellar regime have been reported in
$\sigma$ Ori ($\alpha=0.2\pm0.2$) for the masses  0.004 - 0.19\,M$_\odot$  \citep{damian23}. Earlier studies in the same cluster reported a slope of  $\sim$0.6 for a similar mass range \citep{caballero07,pena12}. However, the work by \citet{damian23}, based on a complete sample of spectroscopically confirmed members, likely provides a more reliable representation of the cluster.
In 25~Ori, \citet{suarez19} report 
a slope $\alpha=0.26\pm0.04$ for the masses between 0.012 and 0.4\,M$_\odot$. With the exception of Lupus~3, all other regions shown in Fig.~\ref{fig:imf_comp} show IMF slope $\alpha \gtrsim$ 0.5 for the masses below 0.2\,M$_\odot$. 
If taken at face value, the scatter in the slope $\alpha$ may appear significant, but it is important to emphasize that the error bars given in Table~\ref{tab:imf_regions} and in most other works reflect only statistical uncertainties and do not account for other potential sources of error or systematics, such as uncertainties in distance and age, the choice of isochrones, or the adopted extinction law. 

\begin{figure}[]
    \centering
    \includegraphics[width=0.45\textwidth]{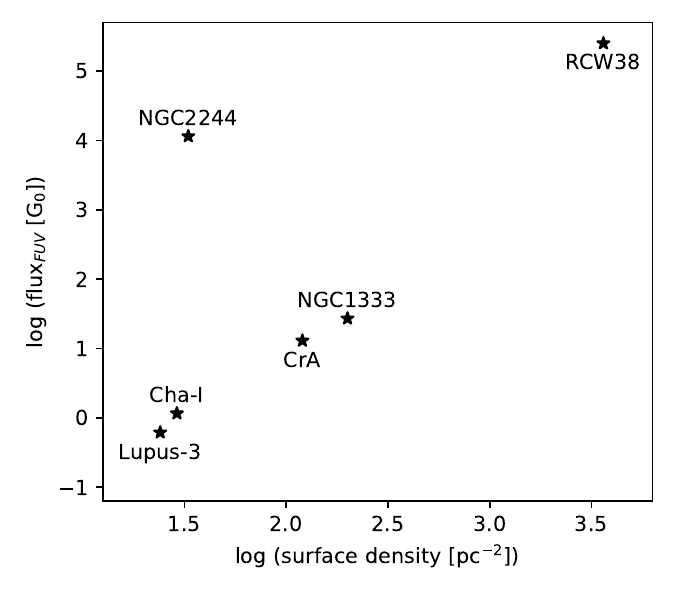}
    \caption{Stellar surface density versus the median FUV flux for the six SFRs and clusters discussed in the text.}
    \label{fig:environment}
\end{figure}

\begin{figure}[]
    \centering
    \includegraphics[width=0.45\textwidth]{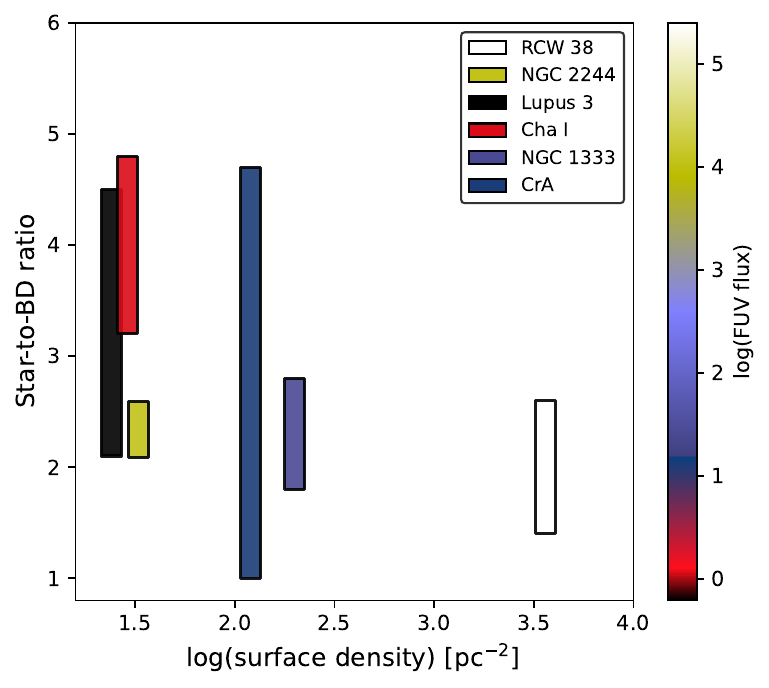}
    \caption{Star-to-brown dwarf ratio in various star-forming regions as a function of stellar surface density. Each rectangle represents a different cluster, height corresponding to the range star-to-brown dwarf ratios; the width is arbitrarily chosen. The color of each rectangle indicates the median FUV flux in the region.}
    \label{fig:sbdratio_regions}
\end{figure}

\begin{figure}[]
    \centering
    \includegraphics[width=0.5\textwidth]{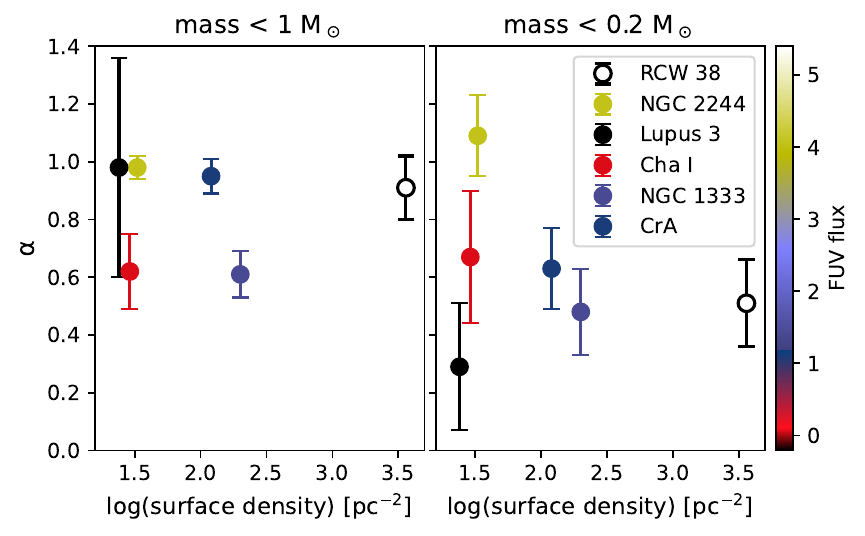}
    \caption{The IMF slope $\alpha$ as a
function of stellar surface density, fitted for the masses $<1\,$M$_\odot$ (left) and $<0.2\,$M$_\odot$ (right). Points are colored by the median FUV flux of each region, similar to Fig.~\ref{fig:sbdratio_regions}.}
    \label{fig:alpha_regions}
\end{figure}

\subsection{Comparison of star-to-BD ratios}
To facilitate a comparison of star-to-BD number ratios, we include the corresponding values in Table~\ref{tab:imf_regions} for the same set of regions used in the IMF analysis. The star-to-BD ratios span a range of approximately 2 to 5. As discussed in \citet{scholz13}, this range reflects various sources of uncertainty in the mass determination, issues related to the membership, and the small sample sizes. It is worth noting that the regions listed in Table~\ref{tab:imf_regions} represent different star-forming environments, which we attempt to characterize using their average stellar density and incident FUV flux (see Fig.~\ref{fig:environment})\footnote{The details on how these values have been obtained are given in Appendices~\ref{sec:density} and \ref{sec:fuv}.}.
The environment of CrA is most similar to that of NGC 1333, as both regions are denser and experience stronger FUV radiation fields compared to the other nearby star-forming regions, Lupus 3 and Cha I. The two massive clusters, NGC 2244 and RCW 38, also occupy distinct positions in this parameter space, both characterized by very strong radiation fields, but representing opposite extremes in terms of stellar density.

Fig.~\ref{fig:sbdratio_regions} shows the star-to-BD ratio for the six regions discussed in this section, plotted as a function of stellar surface density. Each region is represented by a rectangle of arbitrary width, with its color indicating the median FUV flux within the region. Fig.~\ref{fig:alpha_regions} shows a similar plot for the values of the IMF slope $\alpha$, for the fits over two different mass ranges (given in Table~\ref{tab:imf_regions}.
At this point, neither the IMF nor the star-to-BD ratio shows a significant dependence on cluster stellar density or the ionizing flux from the most massive stars. 
A combined effect, where one property might promote BD formation while the other suppresses it, also appears unlikely, as the sample includes regions spanning a range of both parameters. For instance, ChaI has a lower number of BDs per star compared to NGC2244, with the main difference between the two clusters being their FUV radiation levels. Conversely, NGC2244 and NGC1333 exhibit similar star-to-BD ratios, despite the FUV flux in NGC2244 being three orders of magnitude higher. RCW38 also shows a comparable ratio, despite its much higher stellar density relative to both clusters.

Recently, \citet{gupta24} compiled literature values for the star-to-BD ratio across 15 young clusters and SFRs, finding that the ratio decreases with stellar density but shows no clear correlation with incident UV flux.
Based on our analysis, we do not observe a similar decrease in the star-to-BD ratio with stellar density. 
First of all, as noted in their paper, the star-to-BD ratios for some clusters were derived by extrapolating the stellar IMF into the substellar regime. This includes Westerlund~1, which exhibits the lowest ratio in their sample of 15 clusters, and therefore significantly influences the claim of environmental dependence. Additionally, several of the SONYC-derived values are presented with error bars that appear underestimated relative to those in the original publications. For these reasons, we approach the results of \citet{gupta24} with caution.
That said, although our sample is more carefully controlled, it remains limited in size. Furthermore, our analysis of NGC 2244 and RCW 38 is confined to the central regions of the two clusters. Additional observational efforts are essential to reliably evaluate the impact of stellar density on brown dwarf formation, especially in more extreme environments. In parallel, a more uniform treatment of existing datasets would aid in mitigating various sources of systematic uncertainty. This is the primary reason we focus our comparisons on clusters studied within our own work, although even this approach presents challenges due to the evolving nature of the field—such as updates to distances, extinction laws, and theoretical models over the extended period during which the data were collected.

\section{Summary and conclusions}
\label{sec:summary}

We conducted a comprehensive survey of the substellar population in the Corona Australis star-forming region in order to derive its IMF and characterize its low-mass stellar and substellar content. Our approach combines deep optical imaging from Suprime-Cam/Subaru, near-infrared data from the VISTA Hemisphere Survey, and follow-up spectroscopy with KMOS/VLT and FLOYDS/LCO.

The optical spectroscopy of eight kinematically selected candidates using FLOYDS/LCO confirmed them as low-mass stellar members, with spectral types M1 to M5.
In contrast, from a photometric sample of 173 BD candidates identified based on color-magnitude diagrams and evolutionary models, none were confirmed as substellar members of CrA. The KMOS spectroscopic analysis revealed that these objects are predominantly field contaminants, with spectra inconsistent with young, late-type substellar sources.
Nevertheless, these objects help to make stringent constraints on the number low-mass members potentially missing in the census.
Combining our results with the latest census by \citet{esplin22}, we estimate that at most 5 substellar objects may still be missing from the CrA population. 

We derive the CrA IMF over the mass range $0.01$--$1\,M_\odot$, finding a power-law slope of $\alpha = 0.95 \pm 0.06$. 
In the range 0.01 - 0.02\,M$_\odot$, the slope is $\alpha$ = $0.63\pm0.14$, and in the substellar range (0.01 - 0.1\,M$_\odot$), $\alpha$ = $0.33\pm0.19$, indicating the flattening of the IMF in the substellar regime.
The star-to-BD number ratio in CrA is $\sim2$.
For consistent comparisons, we re-derived the IMFs and star-to-BD ratios for Cha~I and Lupus~3 from the SONYC survey, incorporating updated Gaia distances and consistent evolutionary models. For Cha~I, we derive $\alpha = 0.62 \pm 0.13$ over the mass range $0.01-1\,$M$_\odot$, with a star-to-BD ratio of $\sim3.2$--$4.8$. For Lupus~3, we find $\alpha = 0.98 \pm 0.38$ for the mass range $0.02-1\,$M$_\odot$, with a star-to-BD ratio of $\sim2.1$--$4.5$.

Finally, we examined the relationship between the star-to-BD ratio and environmental factors, including stellar surface density and median FUV flux, both consistently derived for the six comparison regions. The stellar surface density varies by over two orders of magnitude, while the FUV fluxes span approximately five orders of magnitude.
Our results show no significant correlation between these properties and the relative abundance of brown dwarfs, suggesting that environmental conditions such as stellar density and ionizing radiation from massive stars do not strongly influence BD formation. 
Our results do not confirm earlier claims of a dependence on stellar surface density \citep{gupta24} and instead support the idea of a largely universal substellar IMF across a range of star-forming environments - at least within the limits of current uncertainties.

\section{Data availability}
Table~\ref{tab:kmos_refuted} is only available in electronic form at the CDS via anonymous ftp to cdsarc.u-strasbg.fr (130.79.128.5) or via http://cdsweb.u-strasbg.fr/cgi-bin/qcat?J/A+A/.

\begin{acknowledgements}
KM and AdB-dV acknowledges support from the Fundação para a Ciência e a Tecnologia (FCT) through the CEEC-individual contract 2022.03809.CEECIND (DOI 10.54499/2022.03809.CEECIND/CP1722/CT0001), and the project 2023.01915.BD (DOI 10.54499/2023.01915.BD). 
KM also acknowledges support from the Scientific Visitor Programme of the European Southern Observatory (ESO) in Chile. BD and AS acknowledge support from the UKRI Science and Technology Facilities Council through grant ST/Y001419/1/. VA-A acknowledges support from the INAF grant 1.05.12.05.03.

\end{acknowledgements}

\bibliographystyle{aa} 
\bibliography{aa55903-25}

\begin{appendix}

\section{KMOS spectra}
\label{sec:kmos_spectra}

Table~\ref{tab:kmos_refuted} provides the coordinates of all objects included in the KMOS follow-up observations. The first eight entries correspond to sources consistent with field M-dwarfs (see Section~\ref{sec:kmos}), while the remaining spectra are inconclusive, i.e. they have been ruled out as young M-type dwarf stars or BDs.

\begin{figure}
    \centering
    \includegraphics[width=0.37\textwidth]{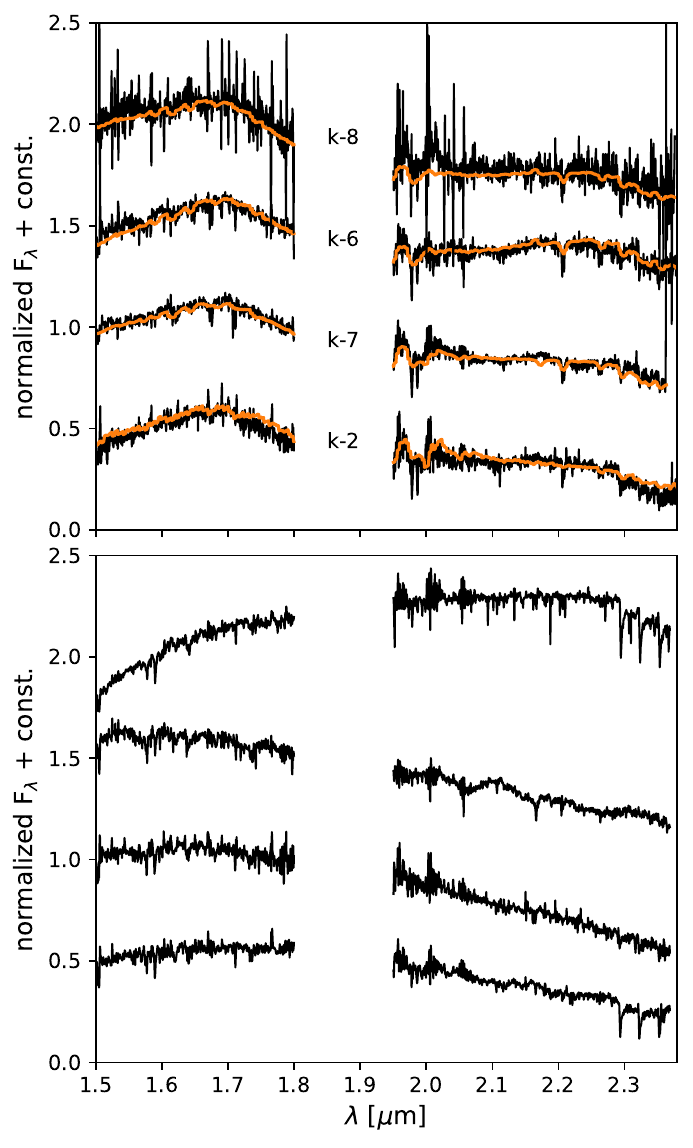}
    \caption{A subset of the KMOS spectra is displayed. The top panel shows several confirmed field M-dwarfs (in black), overlaid with their corresponding best-fit field dwarf templates (in orange). Object IDs are indicated and correspond to those listed in Table~\ref{tab:kmos_refuted}. The bottom panel presents a selection of other spectra from our sample, which can be clearly excluded as young, low-mass M-type stars or brown dwarfs.}
    \label{fig:kmos_spectra}
\end{figure}

In Figure~\ref{fig:kmos_spectra}, we show a subset of the KMOS spectra. The top panel shows several confirmed field M-dwarfs (in black), overlaid with their corresponding best-fit field dwarf templates (in orange). The bottom panel presents a selection of other contaminant spectra.

\begin{table}[!htbp]
	\caption{The objects included in the KMOS follow-up. The first 8 entries are the spectra consistent with being field M-dwarfs, while the remaining objects have inconclusive spectra.}
	\begin{center}
		\begin{tabular}{ccccc}
        \hline \hline
			ID & $\alpha$ & $\delta$ & SpT & A$_V$ (mag) \\
			\hline
			k-1 & 19:02:15.69 & -37:06:02.6 & M6 & 9.0 \\
			k-2 & 19:02:34.05 & -37:00:47.1 & M1 & 9.5 \\
			k-3 & 19:02:41.99 & -36:58:27.9 & M0 & 10.5 \\
			k-4 & 19:02:43.31 & -36:58:37.6 & M5 & 11.5 \\
			k-5 & 19:02:53.51 & -37:04:12.0 & M6 & 10.0 \\
			k-6 & 19:03:13.21 & -37:06:35.9 & M5 & 9.5 \\
			k-7 & 19:03:20.02 & -37:06:37.8 & M4 & 8.0 \\
			k-8 & 19:00:50.52 & -36:59:04.4 & M5 & 6.0 \\
			 & 19:00:20.89 & -36:59:43.9 &  &  \\
			 & 19:00:22.68 & -36:59:52.6 &  &  \\
             \hline
		\end{tabular}
        \tablefoot{The full table is available in the electronic form on CDS.
	}
	\end{center}
	\label{tab:kmos_refuted}
\end{table}

\section{Objects rejected in \citet{galli20}}
\label{sec:galli_rejected}

\begin{figure}[!htbp]
    \centering
    \includegraphics[width=0.45\textwidth]{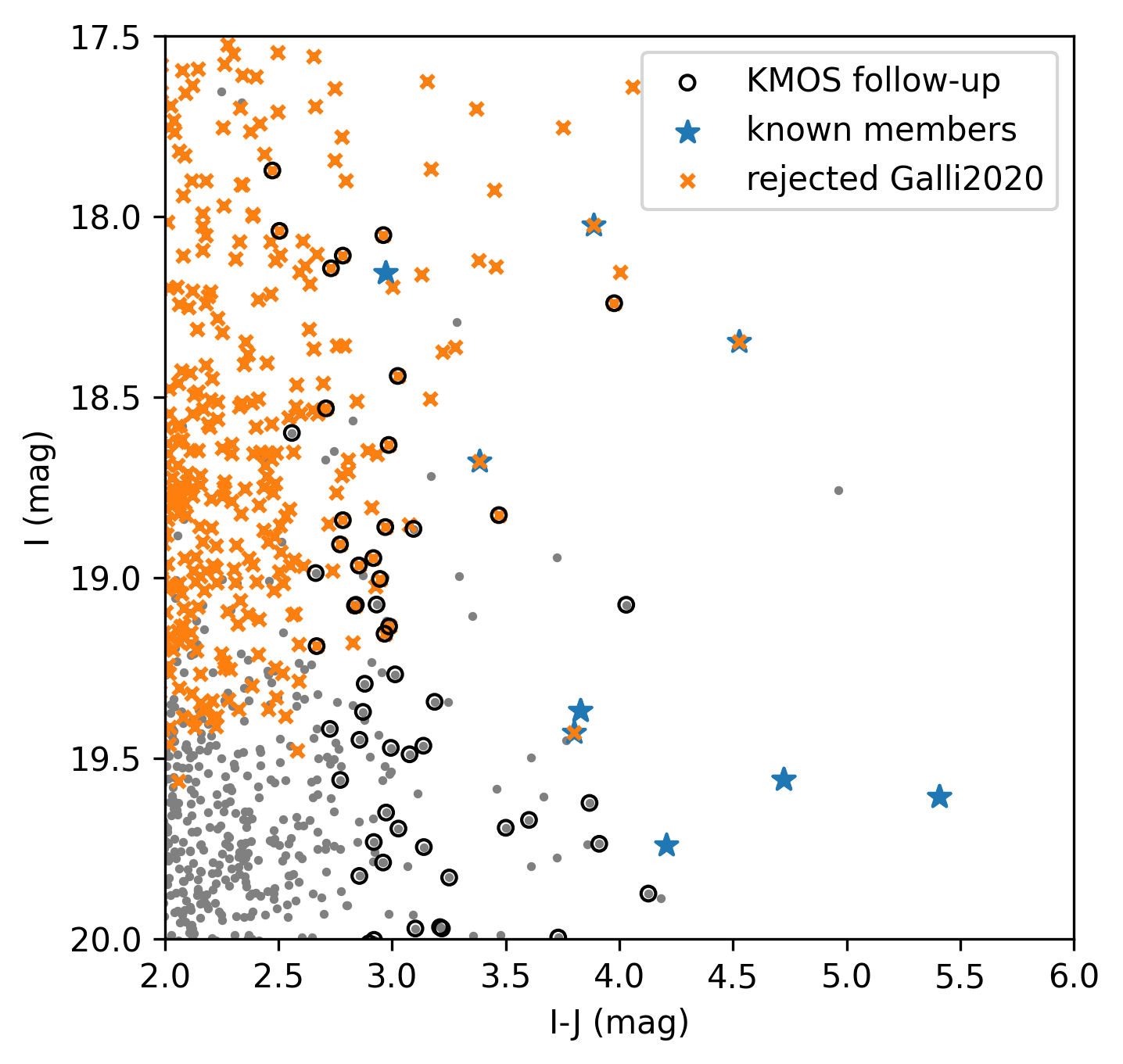}
    \caption{Zoom into the CMD region with an overlap of the KMOS follow-up with Gaia data. The sources in our catalog are shown as grey dots, previously confirmed members from astrometry, spectroscopy, MIR-excess or X-ray emission as blue stars \citep{peterson11,galli20,esplin22}, and objects rejected in \citep{galli20} as orange crosses.
     The sources with spectroscopy with KMOS are marked with black circles.}
    \label{fig:cmd_zoom}
\end{figure}

In Fig.~\ref{fig:cmd_zoom} we show a portion of the CMD shown in Fig.~\ref{fig:cmd}, where exists an overlap between the objects with kinematics from Gaia DR2 studied in \citet{galli20} and our spectroscopic follow-up with KMOS. The symbols are identical to those in Fig.~\ref{fig:cmd}, with an addition of kinematically rejected sources from \citet{galli20}, which are shown as orange points. Three sources originally rejected in this work, were later confirmed as members in \citet{esplin22}.

\section{Updates of SONYC IMFs in Cha I and Lupus 3}
\label{sec:IMF_updated}
In \citet{muzic15}, we published the IMFs in Cha~I and Lupus~3 SFRs, assuming the distances of 160\,pc and 200\,pc, respectively. In the meantime, these distances have been updated thanks to $Gaia$, to 190\,pc for Cha~I \citep{galli21}, and 160\,pc to Lupus~3 \citep{galli20_lupus}. Given the significant discrepancy with the old distances, we recalculated the two IMFs and the corresponding star-to-BD ratios, repeating the identical procedure from the original paper. The newly obtained slopes are consistent with the old ones within the uncertainties. For Cha~I, the IMF slope in the range $0.01 - 1$\,M$_\odot$ changed from $0.78\pm0.08$ to $0.62\pm0.13$, and for Lupus~3 the slope changed from $0.79\pm0.13$
to $0.98\pm0.38$, for the mass range  $0.02 - 1$\,M$_\odot$. 
We note also that the distance to NGC 1333 used in SONYC \citep{scholz12a} is similar to the new distance based on $Gaia$ \citep{ortiz18}, requiring no update.

\begin{figure*}[!htbp]
    \sidecaption
    \includegraphics[width=0.65\textwidth]{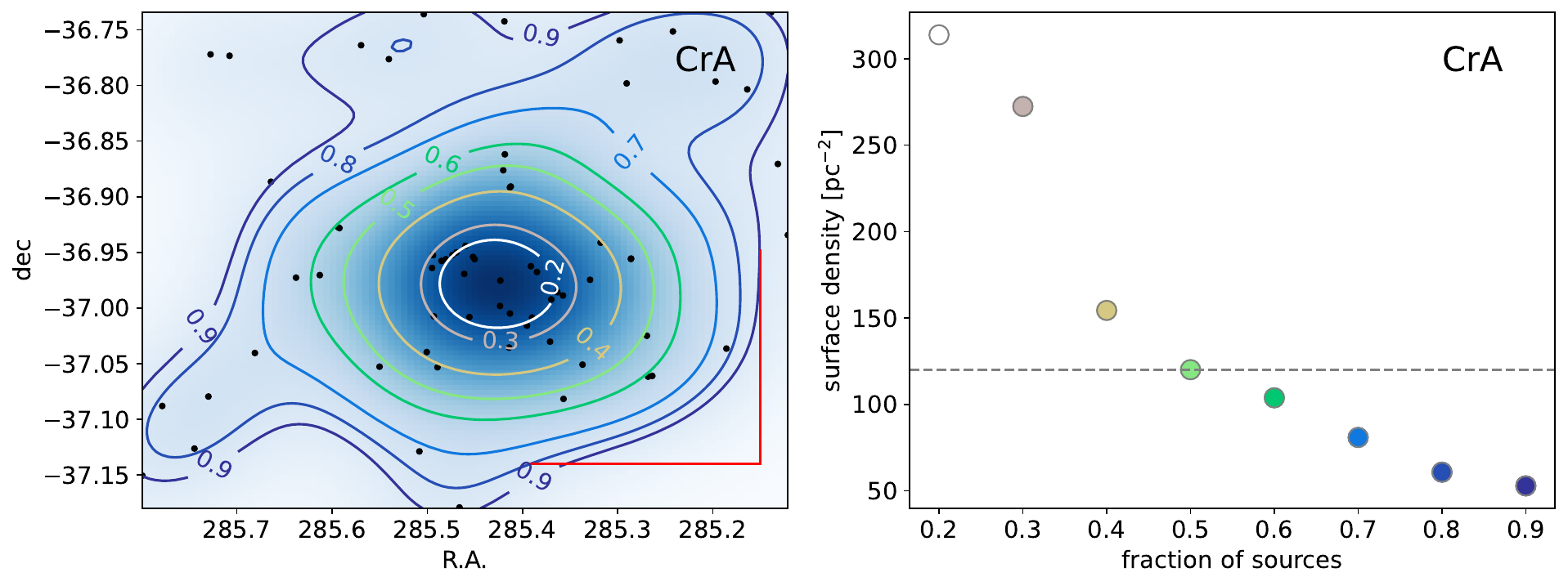}
    \caption{Spatial distribution of members in CrA is shown in the left panel (black dots), overlaid with a two-dimensional kernel density estimate (color map). Contours indicate regions enclosing different percentages of the sources. The corresponding stellar surface densities within each contour are presented in the right panel, with the horizontal dashed line marking the surface density at the 50\% contour level. The red bar in the lower right corner of the left panel represents a scale of 0.5 pc at the distance of CrA.}
    \label{fig:density}
\end{figure*}
\section{Stellar surface densities}
\label{sec:density}

We estimate the surface density of the CrA on-cloud population following the same approach used for other regions in \citet{muzic19}. Specifically, we apply a two-dimensional kernel density estimation (KDE) with a Gaussian kernel to the spatial distribution of cluster members, considering only sources with spectral types earlier than M7. From the resulting KDE map, we define density contours enclosing fixed percentages of members, ranging from 20\% to 90\% (see Fig.\ref{fig:density}). For each contour, we compute the stellar surface density, as shown in the right panel of Fig.\ref{fig:density}. Consistent with \citet{muzic19} and \citet{almendros23}, we adopt the surface density corresponding to the 50\% contour level as a reference value for comparison across regions. For CrA, this value is 120 stars per pc$^2$. 

We also calculated the new values for the stellar surface densities in Cha I and Lupus 3, using the updated distances (see Appendix~\ref{sec:IMF_updated}). The values are listed in Table~\ref{tab:imf_regions}.

\section{Estimate of the FUV flux}
\label{sec:fuv}
Massive stars in a cluster produce a background UV field, which can provide feedback on the star and BD formation process. The amount of radiation cluster stars receive depends on the luminosity of the most massive stars and their distance to other stars. To estimate the FUV field of each cluster, we start by compiling a list of its O and B stars; relevant details and the references are listed at the end of this section. For each spectral subtype, we assume the effective temperature and luminosity from \citet{pecaut13}. The FUV flux is often expressed in terms of the Habing unit G$_0$=$1.6\times10^{-3}$erg\,s$^{-1}$\,cm$^{-2}$, the amount of radiation found for the solar neighbourhood in the wavelength range between 912\AA~and 2400\AA~ \citep{habing68}. 
To estimate a star's FUV luminosity, we scale its total luminosity by the factor calculated using Eq.~4 of \citet{kunimoto21}, specifying the fraction of light emitted in the FUV at a given temperature. The flux received by each cluster star is calculated by dividing the luminosity of each OB source by the square of its distance from the star. Since we only have access to the projected distances in two dimensions, we assume that the distance along the line of sight is half of the projected separation in the plane of the sky.
For each cluster member, we sum the contributions from all ionizing sources. As a measure of the FUV field strength within the cluster, we adopt the median of the fluxes received by all individual stars.

The locations and spectral types of OB stars come from the following sources.
\begin{itemize}
    \item CrA: Four B8-9 stars in the on-cloud population from \citet{esplin22};
    \item Cha I: One B6 and two B9/A0 stars from \citet{luhman07} and \citet{kirk11};
    \item Lupus 3: One B4 member from \citet{kirk11};
    \item NGC\,1333: Two B-type stars (B5 and B8) from \citet{luhman16};
    \item NGC\,2244: Seven O4-O9 and 24 B-type stars from \citet{martins12} and \citet{wang08};
    \item RCW 38: One O5.5 binary (IRS 2; \citealt{derose09} and 17 B-type candidates located in the central part of the cluster from \citet{wolk08}. For these candidates, we assign spectral types according to their bolometric luminosities from \citet{wolk08} following the stellar properties as listed in \citet{pecaut13}.  
\end{itemize}

Recently, \citet{anania25} computed the FUV fluxes for stars bearing discs in 15 regions within $\sim$500\,pc. Fundamentally, their method is similar to ours in that it considers the FUV luminosities of OBA stars and estimates their combined influence on individual stars. The primary difference lies in how they estimate the 3D separations between disc-bearing stars and the massive sources. 
For CrA, they report a median FUV flux of F${\mathrm{FUV,median}}$ = 10$^{+15}_{-6}$, consistent with our estimate. However, for Cha I, our value of 1.2 is lower than the 5$^{+4}_{-1}$ reported by \citet{anania25}.
In contrast, our Cha I estimate aligns with the value reported in \citet{gupta24}, but for three other regions (NGC 1333, NGC 2244, and RCW 38), their estimates appear systematically higher than ours. It is unclear at which distance the fluxes in \citet{gupta24} were calculated, as they only describe how the FUV luminosities were derived.

Given the challenges involved in estimating the FUV fluxes experienced by cluster members, it is perhaps unsurprising that the absolute values reported in different studies show some discrepancies. However, when considered in relative terms, there is broad agreement. For instance, Chamaeleon and Lupus are consistently identified as regions with the lowest median FUV flux, while RCW 38 stands out with fluxes orders of magnitude higher, closely followed by NGC 2244. NGC 1333, in contrast, occupies an intermediate position between these extremes.

\end{appendix}
\end{document}